\documentclass[prc,superscriptaddress,unsortedaddress,twocolumn,showpacs,preprintnumbers,amsmath,amssymb,floatfix]{revtex4}
\usepackage[dvips]{graphicx}
\usepackage{amsmath}
\usepackage{amssymb}
\usepackage{times}
\usepackage{mathrsfs}
\usepackage{multirow}
\usepackage{bm}
\usepackage{wrapft}
\usepackage{color}

\def\ga{\mathrel{\mathpalette\fun >}}
\def\fun#1#2{\lower3.6pt\vbox{\baselineskip0pt\lineskip.9pt
\ialign{$\mathsurround=0pt#1\hfil##\hfil$\crcr#2\crcr\sim\crcr}}}

\newcommand{\beq}{\begin{equation}}
\newcommand{\eeq}{\end{equation}}
\newcommand{\bea}{\begin{eqnarray}}
\newcommand{\eea}{\end{eqnarray}}

\newcommand{\bfi}[1]{\mbox{\boldmath $#1$}}

\newcommand{\vb}{{\bfi b}}

\newcommand{\vK}{{\bfi K}}

\newcommand{\vs}{{\bfi s}}

\newcommand{\vrr}{{\bfi r}}
\newcommand{\vR}{{\bfi R}}

\begin{document}

\pagestyle{empty}


\title{Deformation effect on total reaction cross sections for
neutron-rich Ne-isotopes
}
\author{Kosho Minomo}
\affiliation{Department of Physics, Kyushu University, Fukuoka 812-8581, Japan}

\author{Takenori Sumi}
\affiliation{Department of Physics, Kyushu University, Fukuoka 812-8581, Japan}

\author{Masaaki Kimura}
\affiliation{Creative Research Institution (CRIS), Hokkaido University, Sapporo 001-0021, Japan}

\author{Kazuyuki Ogata}
\affiliation{Department of Physics, Kyushu University, Fukuoka 812-8581, Japan}

\author{Yoshifumi R. Shimizu}
\affiliation{Department of Physics, Kyushu University, Fukuoka 812-8581, Japan}

\author{Masanobu Yahiro}
\affiliation{Department of Physics, Kyushu University, Fukuoka 812-8581, Japan}

\date{\today}

\begin{abstract}
Isotope-dependence of measured reaction cross sections in scattering
of $^{28-32}$Ne isotopes from $^{12}$C target at 240~MeV/nucleon
is analyzed by
the double-folding model with the Melbourne $g$-matrix.
The density of projectile is calculated by the mean-field
model with the deformed Wood-Saxon potential. The deformation is evaluated 
by the antisymmetrized molecular dynamics. The deformation of projectile
enhances calculated reaction cross sections to the measured values.
\end{abstract}


\maketitle

\section{Introduction}

Exploring unstable nuclei is one of the most important subjects in
nuclear physics.
Actually, it was reported that unstable nuclei have exotic properties such as
the halo structure~\cite{Tanihata,Jensen,Jonson}
and the loss of magicity for nuclei in
the so-called ``Island of inversion''.
The term ``Island of inversion'' was first introduced
by Warburton~\cite{Warburton} to
the region of unstable nuclei from $^{30}$Ne to $^{34}$Mg.
In the region, the low excitation energies and the large $B(E2)$ values
of the first excited states suggest
strong deformations~\cite{Mot95,Caurier,Utsuno,Iwas01,Yana03},
which indicates that the $N=20$ magic number is no longer valid.
These novel quantum properties have inspired
extensive experimental and theoretical studies.

Important experimental tools for exploring unstable nuclei are the
reaction cross section $\sigma_{\rm R}$ or the
interaction cross section $\sigma_{\rm I}$
and the nucleon-removal cross section $\sigma_{-n}$ with radioactive
beams~\cite{Tanihata,Jensen,Jonson,Gade};
for the scattering of unstable nuclei,
$\sigma_{\rm I}$ agrees with $\sigma_{\rm R}$ in general, since
projectile excitations to its discrete excited states do not exist.
Very recently, $\sigma_{\rm I}$ was measured
by Takechi {\it et al.}~\cite{Takechi} for $^{28-32}$Ne located near or in
``Island of inversion''.
Furthermore, a halo structure of $^{31}$Ne was reported by the
experiment on the one-neutron removal reaction~\cite{Nakamura}.
This is the heaviest halo nucleus in the present
stage suggested experimentally and also reside
within the region of ``Island of inversion''.

As a useful theoretical tool of analyzing $\sigma_{\rm R}$, we can consider
the microscopic optical potential constructed
by the double-folding model (DFM) with
the $g$-matrix effective nucleon-nucleon (NN) 
interaction~\cite{Satchler-1979,Satchler,M3Y,
Brieva-Rook,JLM,Rikus-von Geramb,CEG,Amos,CEG07},
when the projectile breakup is weak.
For the nucleon-nucleus scattering,
the single-folding model with the $g$-matrix 
well reproduce the data on $\sigma_{\rm R}$ and 
the elastic-scattering cross section
systematically~\cite{Amos}.
For the $^{31}$Ne scattering from
$^{12}$C at 240~MeV/nucleon, the breakup cross section is
at most 1.5\% of $\sigma_{\rm R}$~\cite{ERT}. Hence,
DFM is applicable also for analyses of
measured isotope-dependence of $\sigma_{\rm R}$ in 
the scattering of $^{28-32}$Ne from $^{12}$C target 
at 240~MeV/nucleon~\cite{Takechi}.

In DFM, the $g$-matrix is folded with the projectile and target densities.
If the projectile deforms, the density profile changes;
the surface diffuseness increases because of the elongation.
This gives rise to the effective growth of the root-mean-square (RMS)
radius and eventually the increase of $\sigma_{\rm R}$.
Therefore, the amount of deformation is important.
Nuclei in the island of inversion are spherical
or only weakly deformed in the Skyrme and/or Gogny HF (HFB) calculations;
see, e.g., Refs.~\cite{TFH97,RER02}.
It is even pointed out that the observed large $B(E2;2^+\rightarrow 0^+)$
values can be understood as a large amplitude vibration around
the spherical shape~\cite{YG04}.
In such a situation, the additional correlations by
the angular momentum projection (AMP) often leads to possible deformed shapes;
see Ref.~\cite{RER03} for Ne isotopes.

Recently a systematic investigation employing
the antisymmetrized molecular dynamics (AMD) with the Gogny D1S interaction
has been performed for both even and odd $N$
nuclei in the island of inversion~\cite{Kimura}.
The AMD (with AMP performed) gives rather large deformations,
which is consistent with the AMP-HFB calculations~\cite{RER02,RER03}.
A consistent picture of even and odd isotopes has been obtained,
where the $n$-particle $m$-hole excitations of the Nilsson orbits
play important roles to determine deformed configurations.
Although it is difficult to distinguish the dynamic shape-fluctuation
and static deformation in these light mass nuclei,
one may use the deformed shape suggested by the AMD calculation
to see its effect on $\sigma_{\rm R}$.

In this paper, we analyze the measured isotope-dependence
of $\sigma_{\rm R}$ in scattering of $^{28-32}$Ne isotopes from $^{12}$C 
target at 240~MeV/nucleon, using
DFM with the Melbourne $g$-matrix~\cite{Amos} and
the deformed projectile density suggested by the AMD calculation.

\section{Theoretical framework}

A microscopic optical potential $U$
between a projectile (P) and a target (T) is
constructed with DFM.
The direct and exchange parts, $U_{\rm D}$ and $U_{\rm EX}$, are obtained
by~\cite{DFM-standard-form,DFM-standard-form-2}
\bea
\label{eq:UD}
U_{\rm D}(\vR)&=&\int \rho_{\rm P}(\vrr_{\rm P}) \rho_{\rm T}(\vrr_{\rm T})
            v_{\rm D}(\rho,\vs) d \vrr_{\rm P} d \vrr_{\rm T}, \\
\label{eq:UEX}
U_{\rm EX}(\vR)&=&\int \rho_{\rm P}(\vrr_{\rm P},\vrr_{\rm P}+\vs)
\rho_{\rm T}(\vrr_{\rm T},\vrr_{\rm T}-\vs) \nonumber \\
            &&~~\times v_{\rm EX}(\rho,\vs) \exp{[i\vK(\vR) \cdot \vs/M]}
            d \vrr_{\rm P} d \vrr_{\rm T},~~~~
            \label{U-EX}
\eea
where $\vs=\vrr_{\rm P}-\vrr_{\rm T}+\vR$ for a position vector $\vR$ of P
from T.
The original form of $U_{\rm EX}$ is a non-local function of $\vR$,
but  it has been localized in Eq.~\eqref{U-EX} with the local semi-classical
approximation~\cite{Brieva-Rook}, where $\hbar \vK(\vR)$ is the local momentum
of the scattering considered and
$M=A_{\rm P}A_{\rm T}/(A_{\rm P}+A_{\rm T})$ for the mass number 
$A_{\rm P}$ ($A_{\rm T}$) of P (T).
The validity of this localization is shown in Ref.~\cite{Minomo:2009ds}.
Here, the effective NN interactions, $v_{\rm D}$ and $v_{\rm EX}$,
are assumed to depend on the local density
\bea
 \rho=\rho_{\rm P}(\vrr_{\rm P}+\vs/2)+\rho_{\rm T}(\vrr_{\rm T}-\vs/2)
\eea
at the midpoint of the interacting nucleon pair.

The microscopic potential $U$ is not spherical, 
if one or both of the densities $\rho_{\rm P}$ and 
$\rho_{\rm T}$ are non-spherical. As shown in Ref.~\cite{Satchler-1979}, 
however, the effect is found to be negligible for 
heavy-ion scattering. 
For intermediate incident energies of our interest, 
this can be understood with reasonable approximations. 
For such energies, the rotational motion of deformed P (rotor) is negligible 
compared with the center-of-mass motion of P. 
Hence, the adiabatic approximation is applicable for the rotational motion. 
Under the approximation, P is scattered by $U$ depending 
on not only $\vR$ but also $\Omega$ 
the direction (Euler angles) of the symmetry axis of rotor 
in the space-fixed frame. 
Using the eikonal approximation for 
the center-of-mass motion of P, one can obtain
\bea
   \sigma_{\rm R}=\int d\vb (1-|S|^2)
\label{rotor-sigma}
\eea
with 
\bea
   S=\int \frac{d\Omega}{8\pi^2} 
   \exp \Big[-\frac{i}{\hbar v}\int_{-\infty}^{\infty} dz U(\vR,\Omega)\Big] 
\label{rotor-S}
\eea
for $\vR=(\vb,z)$ and the velocity $v$ of P. Here, we have assumed that 
the ground state of P is a $0^{+}$ state for simplicity. 
The $S$-matrix element $S$ includes back-coupling effects of the 
rotational excitations on the elastic scattering. 
Separating $U$ into the spherical and non-spherical parts, 
$U_0$ and $\Delta U$, we can get 
\bea
S= S_0+ S_0 \int \frac{d\Omega}{8\pi^2} \big( \frac{\delta^2}{2} +\cdots \big)
\eea
with 
\bea
S_0 &=& \exp \Big[-\frac{i}{\hbar v}\int_{-\infty}^{\infty} dz U_0(R)\Big], 
\\
\delta &=& -\frac{i}{\hbar v}\int_{-\infty}^{\infty} dz \Delta U(\vR,\Omega).
\eea
The non-spherical correction to the spherical part $S_0$ 
starts with $\delta^2$, since 
the correction of order $\delta$ vanishes because of the angle average. 
The leading-order correction is significant only at large $b$, since 
$S_0$ vanishes at small $b$ as a result of the strong absorption. 
For large $b$, $\delta$ keeps small because $v$ is large and 
the range of the $z$ integration is small. 
Actually, we confirmed through numerical calculations that 
the leading-order correction to 
$\sigma_{\rm R}$ is 0.01\% for the $^{30}$Ne+$^{12}$C 
scattering at 240~MeV/nucleon; 
note that the error of the eikonal approximation is less than 1\% 
for this scattering. 
Also for the case that the spin of P in its ground state is non-zero, 
it is possible to prove that the leading-order correction to 
$\sigma_{\rm R}$ is of order $\delta^2$. 
Thus, the effect of $\Delta U$ on $\sigma_{\rm R}$ is negligible. 
Therefore, only the spherical part of the density 
is taken in this paper. Detailed discussion 
on this non-spherical effect will be made in the forthcoming paper.

As for $\rho_{\rm T}$, we use the phenomenological $^{12}$C-density
deduced from the electron scattering~\cite{C12-density} by 
unfolding the finite-size effect of the proton charge in the 
standard manner~\cite{Singhal}.
Meanwhile, $\rho_{\rm P}$ is calculated by the mean-field model
with a given average potential or with the self-consistently
determined potential by the Hartree-Fock (HF) method.
No effect of pairing is included for simplicity.
The Ne isotopes (projectiles) under discussions are supposed to be
in the island of inversion (or at its boundary), and expected to be
strongly deformed.
In order to investigate the effect of deformation,
we take a deformed Woods-Saxon (WS) potential~\cite{CDN87},
in which the axially deformed surface $\Sigma(\bm{\beta})$
is specified by the radius,
\begin{equation}
 R(\theta;\bm{\beta}) = R_{0}c_{v}(\bm{\beta})[
 1+{\textstyle \sum_{\lambda}}\beta_{\lambda}Y_{\lambda 0}(\theta)],
\label{eq:surf}
\end{equation}
with the deformation parameters $\bm{\beta} \equiv \{\beta_{\lambda}\}$
and a volume conserving factor $c_{v}(\bm{\beta})$.
The potential value is determined
by replacing the quantity $(r-R_0)$ in a spherical potential
to the distance from the surface $\Sigma(\bm{\beta})$
(with minus sign if the point is inside it).
The Coulomb potential created by charge $(Z-1)e$ distributed
uniformly inside the surface $\Sigma(\bm{\beta})$ in Eq.~\eqref{eq:surf}
is included for protons.
The single-particle eigenstates are calculated by
the (cylindrical) harmonic oscillator basis expansion.
More than twenty oscillator shells are included and
the convergence of the result is carefully checked
to obtain reliable density distributions.
The nucleon density is obtained by
summing up the contributions of occupied Nilsson levels.
The density distribution thus calculated $\rho_{\rm P}^{\rm (in)}(r,\theta)$
is the one in the intrinsic (body-fixed) frame,
and depends on the polar angle $\theta$ from the symmetry axis.
As mentioned above, 
the density in the laboratory frame used in DFM
(Eqs.~\eqref{eq:UD} and~\eqref{eq:UEX}) is obtained by the angle average:
\bea
\rho_{\rm P}(r)=
\frac{1}{2}\int_0^\pi\rho_{\rm P}^{\rm (in)}(r,\theta)\sin{\theta}d\theta.
\eea
We have checked that the angle-averaged
density agrees with high accuracy with the density
calculated by the angular momentum projection from
the Slater determinantial wave function composed of the occupied WS orbits.

No center of mass (CM) correction is included for the calculation of
the density.  We have checked by the spherical Gogny HF calculation
that the CM correction (including the two-body contributions)
to the RMS radius is about 1\% reduction for all the isotopes.
The amount of reduction is smaller than the enhancement caused
by the deformation effect, but is non negligible;
we will return to this point latter.

\section{Results}

We test the accuracy of DFM with the Melbourne $g$-matrix
for $^{12}$C+$^{12}$C scattering at 250.8 MeV/nucleon.
As shown in Table~\ref{table1},
$\sigma_{\rm R}$ calculated with the Melbourne $g$-matrix
is consistent with the experimental data; more precisely,
the latter is slightly smaller than the former by the factor $F=0.982$.
The table also shows the result of the Love-Franey
$t$-matrix nucleon-nucleon interaction in which the nuclear medium effect 
is not included. 
The difference between the two theoretical results
is about 122~mb that corresponds to 16\% of the experimental data. 
Thus, the medium effect is important
at this incident energy.
For the $^{27}$Al +$^{12}$C scattering at 250.7~MeV/nucleon, 
$\sigma_{\rm R}$ calculated with the phenomenological $^{27}$Al 
density~\cite{C12-density} and the normalization factor $F$ 
is 1164~mb, while the experimental value 
is $1159 \pm 14$~mb~\cite{expC12C12}. 
The normalization procedure thus justified is 
applied for the $^{28-32}$Ne +$^{12}$C scattering at 240~MeV/nucleon
analyzed below.


\begin{table}
\caption{
 Reaction cross sections for $^{12}$C+$^{12}$C scattering
at 250.8~MeV/nucleon for two types of effective nucleon-nucleon
interactions.
The cross sections are presented in units of mb.
}
\label{table1}
\begin{center}
\begin{tabular}{cccccc} \hline
   & Exp.~\cite{expC12C12}
   & Love-Franey~\cite{Love-Franey}
   & Melbourne-$g$~\cite{Amos}
   \\ \hline 
   &  782.0 $\pm$ 10
   & 918
   & 796
   \\ \hline 
\end{tabular}
\end{center}
\end{table}

\begin{figure}[htbp]
\begin{center}
 \includegraphics[width=0.4\textwidth,clip]{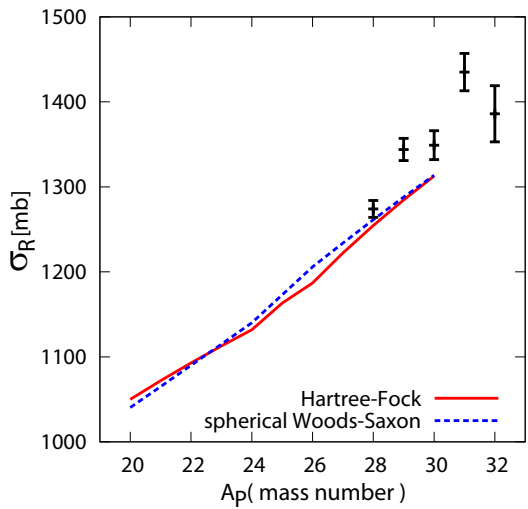}
 \caption{(Color online)
Reaction cross sections for scattering of Ne isotopes from
$^{12}$C at 240~MeV/nucleon.
The results obtained with the calculated density with the Woods-Saxon
potential are denoted by dashed line, and those with the Gogny HF
by solid line.  The spherical shape is imposed.
The nuclei with $A > 30$ are unbound. 
The experimental data are taken from Ref.~\cite{Takechi}. 
 }
 \label{Fig-1}
\end{center}
\end{figure}

\begin{figure}[htbp]
\begin{center}
 \includegraphics[width=0.35\textwidth,clip]{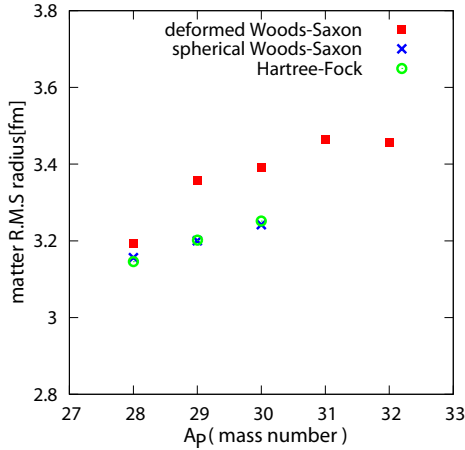}
 \caption{(Color online)
Matter RMS radii for Ne isotopes for 
the spherical WS potential (crosses), the deformed WS potential (squares)
and the Gogny HF (circles). 
 }
 \label{Fig-3}
\end{center}
\end{figure}

As for the parameter set of the WS potential, i.e., the depth, radius and
the diffuseness of the central as well as the spin-orbit potentials,
we employ the one provided recently by R.~Wyss~\cite{WyssPriv};
see Table I of Ref.~\cite{SS09} for the actual values of parameters.
This set is intended to reproduce the spectroscopic properties of
high-spin states from light to heavy deformed nuclei,
e.g., the quadrupole moments and the moments of inertia,
and at the same time the RMS radii crucial for the present analysis.
In order to check that the present WS potential gives reasonable results,
we compare in Fig.~\ref{Fig-1} the reaction cross sections calculated
by using two densities; one obtained by the Gogny D1S HF calculation
and another with the WS potential:
The spherical shape is imposed with the filling approximation
in this calculation.
A good agreement shown in the figure indicates that
the density distributions in the two models are similar,
which is also confirmed by the calculated RMS radii (see Fig.~\ref{Fig-3}).

The reaction cross section is sensitive to
the amount of deformation.
We then employ the deformed shapes suggested by the AMD calculation
to see the effect on $\sigma_{\rm R}$.
As a simple estimate we only include the $Y_{20}$ deformation
in Eq.~\eqref{eq:surf} and the deformation parameter $\beta_2$ in each isotope
is determined to reproduce the calculated ratio of RMS radii
along the long and short axes by AMD;
the resultant values used in the following analyses are
given in Table~\ref{tab:def}.
With these $\beta_2$ values, the Nilsson orbits of the last-odd-neutron
in $^{29}$Ne and $^{31}$Ne are $[200]1/2$ and $[321]3/2$, respectively,
in accordance with the AMD calculation.
Note that the nucleus $^{28}$Ne is at the boundary of
the ``Island of inversion'', and AMD predicts
strong mixing between the states with oblate and prolate deformation.
In the present calculation, we have employed the $\beta_2$ value
of the oblate minimum, which is the main component.

\begin{table}
\caption{
 Deformation parameter $\beta_2$
 used in the calculation of density of Ne isotope
 (those with higher multipoles $\lambda > 2$ are not included).
}
\begin{center}
\begin{tabular}{ccccccc} \hline
 nuclide  &  ~  & $^{28}$Ne      & $^{29}$Ne & $^{30}$Ne & $^{31}$Ne
 & $^{32}$Ne \\ \hline 
 $\beta_2$ &         & $-$0.291        & 0.445     & 0.400     & 0.422
 & 0.335     \\ \hline
\end{tabular}
\end{center}
\label{tab:def}
\end{table}

\begin{figure}[htbp]
\begin{center}
 \includegraphics[width=0.35\textwidth,clip]{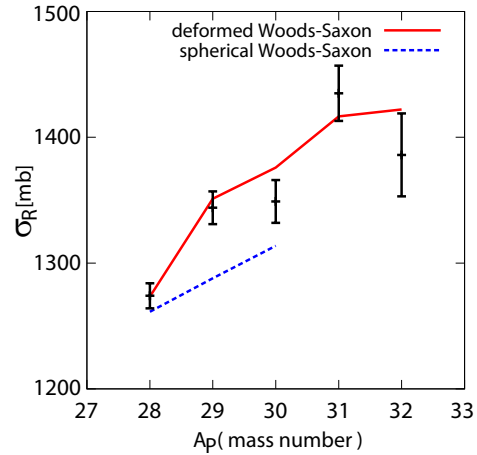}
 \caption{(Color online)
Reaction cross sections for scattering of Ne isotopes from
$^{12}$C at 240~MeV/nucleon. 
The dashed and solid lines represent results of 
the spherical and deformed WS potentials, respectively. 
The experimental data are taken from Ref.~\cite{Takechi}. 
 }
 \label{Fig-2}
\end{center}
\end{figure}

The results of $\sigma_{\rm R}$ including the effect
of quadrupole deformation (see Table~\ref{tab:def})
are shown in Fig.~\ref{Fig-2}.
Compared to the results with the density of the spherical cases,
the effect of deformation
increases the cross section considerably. 
The enhancement makes the calculated cross sections almost 
consistent with the measured cross 
sections for $^{28-32}$Ne, although 
the difference of $\sigma_{\rm R}$ between $^{30}$Ne and $^{31}$Ne
is small in the model calculation compared with the difference deduced 
from the measured cross sections. We will return to this point later. 

\begin{figure}[htbp]
\begin{center}
 \includegraphics[width=0.35\textwidth,clip]{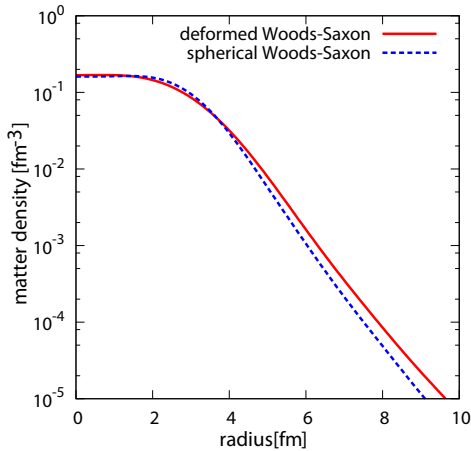}
 \caption{(Color online) The radial dependence
 of the matter density for $^{29}$Ne. 
 The dashed and solid lines show results of the 
 spherical and deformed WS potentials, respectively. 
 }
 \label{Fig-4}
\end{center}
\end{figure}

The increase of $\sigma_{\rm R}$ caused by
the deformation can be rather nicely
understood if one looks into the (matter) RMS radii
$\langle r^2 \rangle^{1/2}_{\rm P}$ shown in Fig.~\ref{Fig-3}.
They are calculated by using the projectile density $\rho_{\rm P}$
based on the spherical and deformed WS potentials.
The increase of $\langle r^2 \rangle^{1/2}_{\rm P}$
in Fig.~\ref{Fig-3} nicely corresponds to that of
$\sigma_{\rm R}$ in Fig.~\ref{Fig-2}, which is reasonable
because of a simple estimate,
\bea
\sigma_{\rm R} \approx \pi
[\langle r^2 \rangle^{1/2}_{\rm P} + \langle r^2 \rangle^{1/2}_{\rm T}]^2 ,
\label{relation}
\eea
where $\langle r^2 \rangle^{1/2}_{\rm T}$ is the RMS radius
for target.
Note that the amount of increase of the RMS radii
from the spherical shape, which is roughly proportional to $\beta_2^2$, 
is only $4-6$\%: It is surprising that such a small effect
is detectable in experimental data.
The present analysis clearly tells us that $\sigma_{\rm R}$
reflects very precise information, and its measurement
is extremely useful to study the nuclear structure of unstable nuclei.
The radial dependence of the matter density is plotted in Fig.~\ref{Fig-4}.
The deformed WS density (solid curve) is enhanced
by the deformation effect from the spherical WS density (dashed curve)
at $r \ga 4$~fm.  The enhancement of the reaction cross section
is caused by that of the density in this tail region.
This is the main reason why we do not directly use the calculated AMD density,
which decreases more rapidly in the tail region
because of the usage of one-range gaussian wave functions.

\section{Discussions}

The enhancement of the reaction cross sections caused
by the deformation effect is conspicuous as shown in Fig.~\ref{Fig-2}.
The enhancement makes the calculated cross sections almost consistent with 
the observed ones for $^{28-32}$Ne; 
more precisely, the calculated cross sections slightly overshoot 
the data for $^{30,32}$Ne, but slightly undershoot the data for $^{31}$Ne. 
In the spherical HF calculation, the CM correction to the RMS radii 
yields 1\% reduction. This leads to 1.1\% reduction 
of $\sigma_{\rm R}$ through relation \eqref{relation}. 
It is very likely that the deformed WS model has almost 
a same amount of $\sigma_{\rm R}$ reduction. 
After this reduction, the calculated cross sections agrees with 
the data for $^{28-30,32}$Ne, but underestimates 
the data by $32\pm22$~mb for $^{31}$Ne. 
Thus, the theoretical results are consistent with the data for  
$^{30}$Ne but not for $^{31}$Ne, because 
the difference of $\sigma_{\rm R}$ between the two nuclei is smaller 
in the model calculation than in the data. 

The difference of $\sigma_{\rm R}$ between 
$^{31}$Ne and $^{30}$Ne 
corresponds to the one-neutron removal cross 
section of $^{31}$Ne, if the breakup cross section 
of $^{31}$Ne is negligible~\cite{ERT}. 
The difference between the observed reaction cross sections is 
86~mb, while the direct measurement on the one-neutron removal 
cross section yields 79~mb~\cite{Nakamura}. 
Thus, the two experimental data are consistent with each other, indicating 
that the breakup cross section is small. 
Meanwhile, the difference of the calculated reaction cross sections between 
the two nuclei is 41~mb and smaller than the experimental results.

As for $^{31}$Ne, the single-particle energies of the last neutron are
about $-2$~MeV in the present deformed WS potential
with $\beta_2$ value given in Table~\ref{tab:def}.
The underestimation of the present value for $^{31}$Ne
may mean that either the depth of the present WS potential is too deep 
or $\beta_2$ is too small. 
For example, compared with the WS potential in Ref.~\cite{Hamamoto10},
the binding energies of relevant Nilsson orbits are about 2 MeV larger 
in the present case, 
though the Nilsson diagrams are very similar to each other.
It turns out that we can obtain good agreements of $\sigma_{\rm R}$
for $^{31}$Ne either by shallowing the potential depth by 
factor $0.943$ or by increasing the deformation up to $\beta_2=0.590$.

In the case of $^{31}$Ne,
its spin-parity and neutron configuration are still under debate.
Our prediction of the last-odd-neutron orbit is $[321]3/2$ with
the single-particle energy $-1.947$~MeV. 
The energy increases to $-0.974$~MeV when the potential is reduced 
by a factor 0.943 to account for the observed central value 
of $\sigma_{\rm R}$, while the last-odd-neutron orbit changes to
$[200]1/2$ and the energy decreases to $-2.803$~MeV when 
$\beta_2$ is increased to $0.590$. 
The measured separation energy of $^{31}$Ne, 
0.29 $\pm$ 1.64 MeV~\cite{Sn}, is more consistent with 
the single-particle energy of the shallower potential rather than 
that of larger $\beta_2$.

It should be mentioned that
the present calculation of $\sigma_{\rm R}$
is not sensitive to the isovector properties, e.g.,  the neutron skin.
Although the matter radii calculated with the present WS and
with the Gogny D1S HF (imposing the spherical shape) perfectly agree
and so do the reaction cross sections (see Fig.~\ref{Fig-1}),
the skin thicknesses in the two calculations are rather different:
e.g., $\langle r^2 \rangle^{1/2}_n -\langle r^2 \rangle^{1/2}_p
\approx 0.67$ and $0.41$~fm with the WS and the Gogny HF, respectively,
in $^{30}$Ne.
Additional information is necessary to probe the property
like the skin thickness.

\section{Summary}

Isotope-dependence of measured reaction cross sections in
scattering of $^{28-32}$Ne isotopes from $^{12}$C target at 240~MeV/nucleon
is analyzed by the double-folding model with the Melbourne $g$-matrix.
The density of projectile is calculated by the mean-field 
model with the deformed Wood-Saxon potential. The deformation
is evaluated by the antisymmetrized molecular dynamics.
The deformation of projectile 
enhances calculated reaction cross sections to the measured values.
The increase of the RMS radii by the deformation is only $4-6$\%, but it is
quite important that such a small effect is detectable 
in the experimental data.
Owing to this effect, the calculated reaction cross sections reproduce
the data for $^{28-30,32}$Ne.  For $^{31}$Ne, however,
the present results still underestimate the measured cross sections.
The underestimation may suggest that the extra weak-binding effect
for neutrons plays an important role particularly for $^{31}$Ne.

\vspace*{5mm}

\section*{Acknowledgements}

The authors thank M. Takechi for providing the numerical 
data and H. Sakurai for useful discussions. 
This work is supported in part by Grant-in-Aid for Scientific Research~(C)
No.~22540285 and 22740169 from Japan Society for the Promotion of Science.
The numerical calculations of this work were performed
on the computing system in Research Institute
for Information Technology of Kyushu University.


\end{document}